\documentclass{ws-p8-50x6-00}

\newcommand{\beq}{\begin{equation}}
\newcommand{\eeq}[1]{\label{#1} \end{equation}}

\begin{document}

\title{Dual Properties of the Structure Functions}

\author{\underline{L.L. Jenkovszky}}\address{Bogolyubov
Institute for Theoretical Physics,
Academy of Sciences of Ukraine\\
01143 Kiev, Ukraine; E-mail:jenk@gluk.org} \author{V.K. Magas}\address{Center for Physics of Fundamental
Interactions (CFIF), Instituto Superior Tecnico\\
Av. Rovisco Pais, 1049-001 Lisbon, Portugal;\\
Bogolyubov Institute for Theoretical Physics,
Academy of Sciences of Ukraine\\
01143 Kiev, Ukraine; E-mail: vladimir@cfif.ist.utl.pt}

\maketitle

\abstracts{By using the concept of duality between direct channel
resonances and Regge exchanges we relate the small- and large-$x$
behavior of the structure functions. We show that even a single
resonance exhibits Bjorken scaling at large $Q^2$.}

In a number of recent papers \cite{JMP,FJM,JM,JM2,JKM} we
have  suggested to apply the concept of resonance-reggeon duality
to study the relation between the small- and large-$x$ behavior of
the nucleon structure functions.

We use standard notations for the cross
section and structure function:
\beq
\sigma^{\gamma^* p}={4\pi^2\alpha(1+4m^2x^2/Q^2
)\over{Q^2(1-x)}}{F_2(x,Q^2)\over{1+R(x,Q^2)}},
\eeq{eq1}
where $\alpha$ is the fine structure constant, $Q^2$ is minus the
squared four momentum transfer, $x$ is the Bjorken variable and $s$ is the squared
center of mass energy of the $\gamma^*p$ system, obeying the
relation
\beq
s=Q^2(1-x)/x+m_p^2,
\eeq{eq2}
where $m_p$ is the
proton mass and $R(x,Q^2)= \sigma_L(x,Q^2)/\sigma_T(x,Q^2)$. For
the sake of simplicity we set $R=0,$ which is a reasonable
approximation.

We use the norm where
\beq
\sigma_T^{\gamma^*}(s,t,Q^2)=Im\ A(s,t,Q^2).
\eeq{eq3}
According to the two-component duality
picture both the scattering amplitude $A$ and the structure
function $F_2$ are sums of a diffractive and non-diffractive
terms. At high energies both terms are Regge-behaved. In $\gamma^*
p$ scattering only positive signature exchange are allowed. The
dominant ones are the Pomeron and the $f$ Reggeon, respectively.
The relevant scattering amplitude is (remember that here $t=0$)
\beq
A_k(s,Q^2)=i\beta_k(Q^2)\Bigl(-i{s\over{s_k}}\Bigr)^{\alpha_k(0)-1},
\eeq{eq4}
where $\alpha$ and $\beta$ are the Regge trajectory and
residue and $k$ stand for the Pomeron or Reggeon. As usual, the
residue will be chosen such as to satisfy approximate Bjorken
scaling for the structure function.

The invariant dual on-shell scattering
amplitude dual amplitude with Mandelstam analyticity (DAMA)
applicable  both to the diffractive and non-diffractive components
reads
\beq D(s,t)=\int_0^1 {dz \biggl({z \over g}
\biggr)^{-\alpha(s')-1} \biggl({1-z \over
g}\biggr)^{-\alpha(t')-1}},
\eeq{eq21}
where $s'=s(1-z), \ \
t'=tz, \ \ g$ is a parameter, $g>1$, and $s, \ \ t$ are the
Mandelstam variables.

For $s\rightarrow\infty$ and fixed $t$ it has the following Regge
asymptotic behavior
\beq
D(s,t)\approx\sqrt{{2\pi\over{\alpha_t(0)}}}g^{1+a+ib}\Biggl({s\alpha'(0)g\ln
g\over{\alpha_t(0)}}\Biggl)^{\alpha_t(0)-1},
\eeq{eq22}
where
$a=Re\ \alpha\Bigl({\alpha_t(0)\over{\alpha'(0)\ln g}}\Bigr)$ and
$b=Im\ \alpha\Bigl({\alpha_t(0) \over{\alpha'(0)\ln g}}\Bigr)$.

The pole structure of DAMA is similar to that of the Veneziano
model except that multiple poles may appear at daughter levels.
The pole term  is a generalization of the Breit-Wigner formula,
comprising a whole sequence of resonances lying on a complex
trajectory $\alpha(s)$. Such a "reggeized" Breit-Wigner formula
has little practical use in the case of linear trajectories,
resulting in an infinite sequence of poles, but it becomes a
powerful tool if complex trajectories with a limited real part and
hence a restricted number of resonances are used. It appears that
a small number of resonances is sufficient to saturate the direct
channel.

Contrary to the Veneziano model, DAMA  does not only allow but
rather requires the use of nonlinear complex trajectories
providing, in particular, for the imaginary part of the amplitude,
resonances widths and resulting in a finite number of those. More
specifically, the asymptotic rise of the trajectories in DAMA is
limited by the condition
$ |{\alpha(s)\over{\sqrt s\ln
s}}|\leq const, \ \ s\rightarrow\infty.$

Our main idea is the introduction of the  $Q^2$-dependence in the
dual model by matching its Regge asymptotic behavior and pole
structure to standard forms, known from the literature. The point
is that the correct identification of this $Q^2$-dependence in a
single asymptotic limit of the dual amplitude will extend it to
the rest of the kinematical regions. We have two ways to do so:
A) Combine  Regge behavior and Bjorken scaling limits of the
structure functions (or $Q^2$-dependent $\gamma^*p$ cross
sections); B) Introduce properly $Q^2$ dependence in the
resonance region.
They should match if the procedure is correct and the dual
amplitude should take care of any further inter- or
extrapolation.

It is obvious  that asymptotic Regge and scaling behavior require
the residue to fall like $\sim(Q^2)^{-\alpha_k(0)+1}$. Actually,
it could be more involved if we require the correct
$Q^2\rightarrow 0$ limit to be respected and the observed scaling
violation (the "HERA effect") to be included. In combining Regge
asymptotic behavior with (approximate) Bjorken scaling, one can
proceed basically in the following way -- keep explicitly a
scaling factor $x^{\Delta}$ (to be broken by some $Q^2$-dependence
"properly" taken into account).
\beq F_2(x,Q^2)\sim
x^{-\Delta(Q^2)}\Bigl({Q^2\over{Q^2+Q_0^2}}\Bigr)^{1+\Delta(Q^2)},
\eeq{eq41}
where $\Delta(Q^2)=\alpha_t(0)-1$ may be a constant, in
particular.

Note that since the Regge asymptotic of the Veneziano model is
$\sim(-\alpha' s)^{\alpha(t)-1},$ the only way to incorporate
there $Q^2-$ dependence is through the slope $\alpha'$, i.e. by
making the trajectories $Q^2-$ dependent, thus violating Regge
factorization \cite{JMP,FJM}. $Q^2$-dependent intercepts were used earlier in a
different context, namely to cope with the observed "hardening" of
small-$x$ physics with increasing $Q^2$ (Bjorken scaling
violation). Although we do not exclude this possibility, we study
here a different option, that by introducing scaling violation in
the residue rather than in the trajectory.

>From the explicit Regge asymptotic form of DAMA and neglecting the
logarithmic dependence of $g$  we make the following
identification
\beq g(Q^2)^{\alpha_t(0)+a}=\left(Q_{lim}^2
\over{Q^2+Q_0^2}\right)^{\alpha_t(0)}
\eeq.
One may notice that
the above equation is, in fact, a transcendent one with respect to
$g$ ($a=a(g)$). Another point to mention is that this equation does not work
in all range of $Q^2$, since for $Q^2$ close to $Q_{lim}^2$ $g$
may get smaller than $1$, which is unacceptable in DAMA. For large
$Q^2$ the $Q^2$-dependence of the $\log g$ and
$b=b(Q^2)$ in eq.(\ref{eq22}) can not be neglected; it might contribute to scaling
violation.

Let us now consider the extreme case of a single resonance
contribution.
A resonance pole in DAMA contributes with (if we avoid the peaks
on daughter trajectories \cite{D})
$$A(s,t)=g^{n+\alpha_t(0)} {C_n\over{n-\alpha(s)}}.$$
At the resonance $s=s_R$ one has $Re\ \alpha(s_R)=n$ and
${Q^2(1-x)\over x}=s_R-m^2$, hence
$$
F_2(x,Q^2)={Q^2(1-x)\over{4\pi^2\alpha\Bigl(1+{4m^2x^2\over
{Q^2}}\Bigr)}} {C_n\over {Im\ \alpha(s_R)}}g(Q^2)^{n+\alpha_t(0)}.
$$
As $x\rightarrow 1$
$Q^2\approx{s_R-m^2\over{1-x}}\rightarrow
\infty$ and
$$F_2(x,Q^2)\sim g\Bigl({s_R-m^2\over{1-x}}\Bigr)^{n+\alpha_t(0)}.$$
By using the approximate solution
$
g(Q^2)\approx\left({Q^2_{lim}/
Q^2}\right)^{\alpha_t(0)\over{\alpha_t(0)+a}},
$
where $a$ is a slowly varying function of $g$, we get for $x$ near $1$
$$F_2(x,Q^2)\sim (1-x)^{\alpha_t(0)(n+\alpha_t(0))\over{\alpha_t(0)+a}},
$$
where the limits for $x$ are defined by
$Q_0^2\ll{s_R-m^2\over{1-x}}\leq Q^2_{lim}.$
We recognize a typical large-$x$ scaling behavior $(1-x)^N$ with
the power $N$ (counting the quarks in the reaction)  depending
basically on the intercept of the $t$-channel trajectory.

The main conclusions from our analysis are that:
A) $Q^2$-dependence at low- and high-x (or high- and low-s) has
the same origin;
B) a single low energy resonance can produce a smooth
scaling-like curve in the structure function (parton-hadron
duality).

{\bf Acknowledgments:\ \ }
We thank E. Predazzi and B.V. Struminsky for
discussions. Both of us acknowledge the support by INTAS, Grant
00-00366. The work of L.J. was supported also by INTAS, Grant
97-1696, and CRDF, Grant UP1-2119.

\end{document}